\pdfoutput=1
\documentclass[11pt]{article}

\usepackage[preprint]{acl}

\usepackage{times}
\usepackage{latexsym}

\usepackage[T1]{fontenc}
\usepackage[utf8]{inputenc}

\usepackage{microtype}

\usepackage{inconsolata}

\usepackage{graphicx}

\usepackage{amsmath}
\usepackage{multirow}
\usepackage{booktabs}

\usepackage{amsfonts}
\usepackage{makecell}
\usepackage{subfigure}
\usepackage{enumitem}

\usepackage{tcolorbox}  % 文本加框
\usepackage{tikz}
\tcbuselibrary{breakable}   % 配合文本加框，使得框可以跨页

\title{SyNeg: LLM-Driven Synthetic Hard-Negatives for Dense Retrieval}

\author{Xiaopeng Li$^{\clubsuit}$, Xiangyang Li$^{\diamondsuit}$, Hao Zhang$^{\diamondsuit}$ Zhaocheng Du$^{\diamondsuit}$ \\ \textbf{Pengyue Jia}$^{\clubsuit}$, \textbf{Yichao Wang}$^{\diamondsuit}$, \textbf{Xiangyu Zhao}$^{\clubsuit}$, \textbf{Huifeng Guo}$^{\diamondsuit}$, \textbf{Ruiming Tang}$^{\diamondsuit}$\\
$^{\clubsuit}$City University of Hong Kong, $^{\diamondsuit}$Huawei Noah’s Ark Lab\\
\texttt{xiaopli2-c@my.cityu.edu.hk}, \texttt{\{lixiangyang34, zhang.hao3, zhaochengdu\}@huawei.com} }

\begin{document}
\maketitle
\begin{abstract}
The performance of Dense retrieval (DR) is significantly influenced by the quality of negative sampling. Traditional DR methods primarily depend on naive negative sampling techniques or on mining hard negatives through external retriever and meticulously crafted strategies. However, naive negative sampling often fails to adequately capture the accurate boundaries between positive and negative samples, whereas existing hard negative sampling methods are prone to false negatives, resulting in performance degradation and training instability. Recent advancements in large language models (LLMs) offer an innovative solution to these challenges by generating contextually rich and diverse negative samples. In this work, we present a framework that harnesses LLMs to synthesize high-quality hard negative samples. We first devise a \textit{multi-attribute self-reflection prompting strategy} to direct LLMs in hard negative sample generation. Then, we implement a \textit{hybrid sampling strategy} that integrates these synthetic negatives with traditionally retrieved negatives, thereby stabilizing the training process and improving retrieval performance. Extensive experiments on five benchmark datasets demonstrate the efficacy of our approach, and code is also publicly available\footnote{The code repository will be available after publication.}.
\end{abstract}

\section{Introduction}

Dense retrieval (DR) is extensively utilized in natural language processing (NLP) and information retrieval (IR) tasks, such as conversational search~\cite{yu2021few}, web search~\cite{craswell2020overview}, question answering~\cite{guu2020retrieval}, etc. Typically, DR models are used to encode queries and documents into text embeddings, 
enabling efficient retrieval of relevant documents from large-scale corpora using similarity measures like inner product or cosine similarity~\cite{chen2024bge,wang2023improving}. It is also key to retrieval-augmented generation (RAG)~\cite{gao2023retrieval}, empowering LLMs to access dynamic external knowledge without modifying their parameters. Clearly, the effectiveness of dense retrieval is crucial for enhancing the performance of IR applications.

Existing DR models~\cite{sun2022reduce} predominantly employ contrastive learning with negative sampling to improve model's discrimination of relevant documents from numerous irrelevant ones. Common techniques like random negative sampling~\cite{huang2020embedding} and in-batch negatives~\cite{karpukhin2020dense} are simplistic, failing to differentiate subtle semantic variances between positive and plausible yet irrelevant samples, \textit{i.e.}, hard negatives. To address this limitation, several advanced sample methods are proposed, such as hard negative mining via external retrievers~\cite{xiong2020approximate,sun2022reduce}, dynamic negative sample updating through iterative training~\cite{zhan2021optimizing}, ambiguous negative sampling~\cite{zhou2022simans}, and quasi-triangular principle-based sampling~\cite{yang2024trisampler}, etc.

Although effective, these approaches have notable drawbacks. First, the quality of sampled negatives is constrained by the auxiliary models, often introducing noise and numerous false negatives, which may degrade model performance. Second, these methods mine negative samples solely from the provided corpus, leading to poor diversity and uneven distribution. Additionally, due to corpus limitations, certain queries may lack high-quality negative samples. While manual construction of high-quality data is feasible, it is labor-intensive and impractical for large-scale datasets.

LLMs for data synthesis~\cite{ma2023pre, wang2023improving} offer an unprecedented opportunity to address these challenges. Given their extensive training on diverse corpora and remarkable capabilities in semantic understanding, language generation, and world knowledge, employing LLMs to generate negative samples provides multiple advantages. Firstly, the quality of negative samples is no longer limited by auxiliary models, as LLMs can autonomously generate negatives based on their understanding of queries and positive samples, thereby reducing the occurrence of false negatives. 
Secondly, the generation of negative samples by LLMs paves the way for creating out-of-distribution (OOD) samples, effectively broadening the corpus and enriching the dataset. This approach is especially useful for queries with sparse negative samples and allows for large-scale production with minimal human intervention.
However, utilizing LLMs to generate high-quality negatives and enhance DR model performance is not easy, two pivotal issues need to be addresses: (1) \textit{How to effectively design prompts for LLMs to ensure the generation of valid negative samples with accurate and diverse distributions?} (2) \textit{How to seamlessly integrate the generated negatives into the training process to elevate the performance of DR models?}

In this work, we propose a simple yet effective framework \textbf{SyNeg} for synthesizing the high-quality negative samples and integrating them to bolster retrieval efficacy. During the negative sample generation phase, we introduce a multi-attribute, self-reflection prompting strategy. This strategy ensures the generation of high-quality and diverse synthetic documents by precisely defining the negative samples and specifying key attributes. Besides, we also incorporate the CoT-based self-reflection mechanism~\cite{wei2022chain} within the generative process to further assure the quality and diversity. Additionally, we devise a hybrid strategy that mixes synthetic negatives with retrieved negatives at the instance level, stabilizing the training process and improving accuracy. Our main contributions are summarized as follows:
\begin{itemize}[leftmargin=*]
    \item We introduce a multi-attribute self-reflection strategy designed to adeptly elicit high-quality hard negatives from LLMs for applications in dense retrieval.
    \item We propose a hybrid strategy that combines generative and retrieved negatives at the instance level, aiming to stabilize and improve the training process.
    \item Experiments on five public datasets in BEIR validate the superior performance of our approach across various dense retrieval models, additionally outperforming existing negative sampling techniques.
\end{itemize}
\section{Theoretical Analysis}\label{sec:preliminary}
This section first introduces dense retrieval and then mathematically proves the importance of high-quality negative samples in training, demonstrating the necessity of our approach, SyNeg.

\subsection{Dense Retrieval}
Given a query $q$, the task of dense retrieval is to seek the most relevant documents from the document corpus. Generally, dual-encoder~\cite{reimers2019sentence} is used to encode query $q$ and doc $d$ into $k$-dimensional embeddings $\boldsymbol{h}_{q}$ and $\boldsymbol{h}_{d}$, respectively, then the relevance is calculated using similarity functions $f$, denoted as:
\begin{equation}\label{equ:similarity} 
    s(q,d) = f (\boldsymbol{h}_{q;\theta}, \boldsymbol{h}_{d;\theta})
\end{equation}
the training objective function of dense retrieval could be represented as:
\begin{equation}\label{equ:loss function} 
    \mathcal{L} = \sum_q\sum_{d^+\in\mathcal{D}^+ }\sum_{d^-\in\mathcal{D}^-} l (s(q,d^+),s(q,d^-) )
\end{equation}
where $l(\cdot)$ denotes the loss function. For simplicity, we adopt the pairwise loss~\cite{ding2020rocketqa} for furthur analysis, denotes as:
\begin{align}
     l (s(q,d^+),s(q,d^-) ) = \tilde{l}(d^+, d^-)= \mathbf{1}_{s(q,d^+) < s(q,d^-)}
\end{align}
here, $\mathbf{1}_{c}$ is an indicator function that equals 1 if the condition $c$ is met, otherwise 0. Thus, we can derive the relationship between loss and the ranking position $\pi(d^+)$ of positive doc $d^+$ as follows:
\begin{equation}\label{equ:rank pos v.s. loss}
    \pi(d^+) = \delta(d^+) + 1 + \sum_{d^- \in \mathcal{D}^-} \tilde{l}(d^+, d^-)
\end{equation}
here, $\delta(d^+)$ represents the positive documents ranked higher than $d^+$, and $\mathcal{D}^-$ is the collection of negative documents.

\subsection{Analysis of Negative Sampling}
Currently, the most widely used random sampling is hard-negative sampling~\cite{xiong2020approximate,zhou2022simans,yang2024trisampler}, which sample the top-$N$ documents as negatives,where $N$ denotes the number of hard negatives, and $N_q$ denote the ranking position of the $N$-th negative document. We can derive that:
\begin{equation} \label{equ:ranking position}
\sum_{d^- \in D^-} \mathbf{1}_{\pi(d^-) \leq N_q} = N
\end{equation}
thus, the training objective with top-$N$ negatives can be derived as:
\begin{equation}\label{equ:top-k neg}
\begin{aligned}
\mathcal{L} &= \sum_{q}\sum_{d^+\in\mathcal{D}^+}\sum_{d^- \in \mathcal{D}^-} \mathbf{1}_{\pi(d^-) \leq N_q} \cdot \tilde{l}(d^+, d^-) \\
&= \sum_{q} \sum_{d^+ \in \mathcal{D}^+} \min(\pi(d^+) - \delta(d^+) - 1, N)
\end{aligned}
\end{equation}
clearly, loss $\mathcal{L}$ is bounded by $N$, resulting in $\mathcal{L}$ is not affected by specific query $q$ where $\pi(d^+)$ might be very large. Compared with random negatives, using top-$N$ negatives could be more robust in retrieval performance, that is also discussed by~\citet{zhan2021optimizing}. However, practically, we cannot directly obtain the top-$N$ negatives but must use retriever to pre-retrieve the top hard negatives. We describe this sampling strategy as follows:
\begin{equation}\label{equ:pre-retrieved obj}
\begin{aligned}
\sum_q \sum_{d^+ \in \mathcal{D}^+} \sum_{d^- \in \mathcal{D}^-} \mathbf{1}_{d^- \in \mathcal{D}_g^-} \cdot \tilde{l}(d^+, d^-)
\end{aligned}
\end{equation}
here, $\mathcal{D}_g^-$ denotes the pre-retrieved documents by retriever $g$ and $\mathcal{D}^-_g \subseteq \mathcal{D}^-$. Following previous research~\cite{zhan2021optimizing}, we define $\phi(\mathcal{D}_g^-)$ as the quality of set $\mathcal{D}_g^-$, using the highest
ranking position of the sample in $\mathcal{D}_g^-$, shown as follows:
\begin{equation}\label{equ:quality}
\phi(\mathcal{D}_g^-) = \min_{d^- \in \mathcal{D}_g^-} \pi(d^-) \in [1, \pi(\mathcal{D}^-_g) ]
\end{equation}
we could see that $\phi(\mathcal{D}_g^-)$ is bounded by $\pi(\mathcal{D}^-_g)$, indicating that the upper bound is closely related to the chosen retriever $g$. The more powerful the retriever, the tighter the bound on $\phi(\mathcal{D}_g^-)$, as it retrieves more high-ranking negative samples. Thus, we can reformulate Equation~\eqref{equ:pre-retrieved obj} and use MRR (Mean Reciprocal Rank) as an evaluation metric. The infimum of MRR metric can be written as:
\begin{equation}\label{equ:pre-retrieved inf}
\begin{aligned}
\inf \text{MRR} &= E_q \frac{1}{ \phi(\mathcal{D}_g^-) - |\mathcal{D^+}|}
\end{aligned}
\end{equation}
here, $\inf$ denotes the infimum, and we can see that the infimum is associated with $\phi(\mathcal{D}_g^-)$. 

In this paper, we propose a pipeline that integrates LLM-generated synthetic negatives with the set of retrieved hard negatives, denoted as $\widetilde{\mathcal{D}}^-_g$. Since $\widetilde{\mathcal{D}}^-_g$ includes high-quality negatives generated by LLMs, which rank higher in top positions, thus for quality comparison, we can conclude that:
\begin{equation}\label{equ:inf comparison}
\begin{aligned}
\phi ( \widetilde{\mathcal{D}}^-_g) < \phi(\mathcal{D}_g^-) 
\end{aligned}
\end{equation}
combine with Equations~\eqref{equ:pre-retrieved inf}, we deduce that the presence of synthetic negatives can elevate the infimum, thereby enhancing the performance of MRR. Thus from the above, we mathematically prove the validity of our approach, and the specific descriptions are depicted in Section~\ref{sec:Methodology} and the results in Section~\ref{sec:Experiment} also confirm our conclusions.

\begin{figure*}[t]
    \centering
    \includegraphics[width=0.85\linewidth]{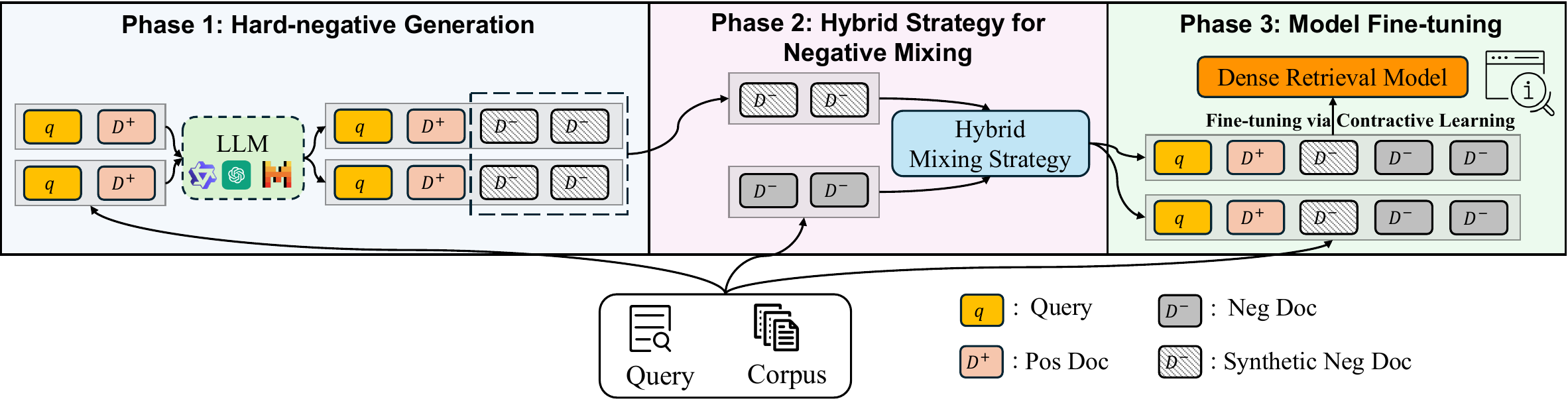}
    \vspace{-2mm}
    \caption{Overview of SyNeg.}
    \vspace{-4mm}
    \label{fig:framework}
\end{figure*}

\begin{figure}[h]
    \centering
    \vspace{-2mm}
    \includegraphics[width=0.9\linewidth]{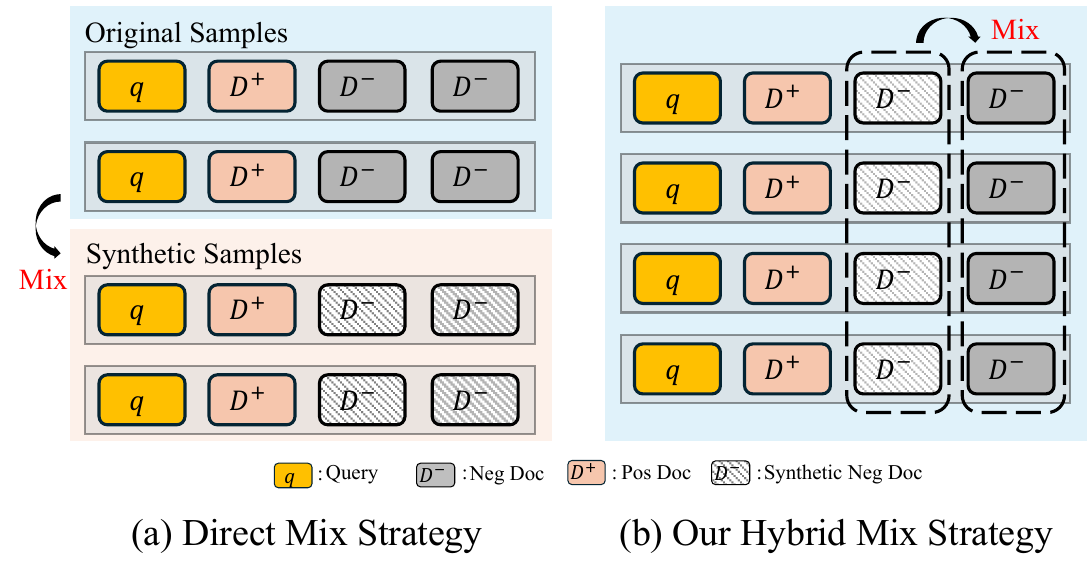}
    \vspace{-2mm}
    \caption{Different mixing strategy comparison.}
    \vspace{-3mm}
    \label{fig:mix strategy}
\end{figure}

\section{Methodology}\label{sec:Methodology}

In this section, we will first provide an overview framework, followed by a detailed description of each subsection and the different phases of our methodology.

\subsection{Overview}

Our framework SyNeg is illustrated in Figure~\ref{fig:framework}, including three distinct phases. Firstly, we introduce a multi-attribute, self-reflection strategy to prompt LLMs for the generation of high-quality negatives. This approach defines key attributes relevant to the negatives, such as knowledge domain and complexity level, etc., and combines a self-reflection mechanism that enables LLMs to assess the quality, diversity, and length of the generated negatives comprehensively. In the subsequent phase, we advocate a hybrid strategy that mixes synthetic hard negatives with retrieved negatives at the instance level, aiming to stabilize and enhance the robustness of the training process. 
The final phase involves fine-tuning, where we use the integrated data from previous phases and apply contrastive learning to optimize the dense retrieval models. The following sections will detail these phases.

\subsection{Hard-negative Generation} \label{sec:hard-neg generation}

The use of LLMs to generate high-quality hard negatives is central to our methodology. Previous studies~\cite{dai2022promptagator, almeida2024exploring, lee2024gecko} have explored generating queries with LLMs, and \citet{wang2023improving} generate instances that combine queries, positive documents, and negatives. In contrast, our work solely focuses on generating negatives to enhance retrieval performance as hard negative sampling significantly boosts the effectiveness within the pipeline of dense retrieval~\cite{karpukhin2020dense, zhan2021optimizing, yang2024trisampler}. Detailed prompt and cases are provided in Appendix~\ref{sec:Appendix Prompt} and~\ref{sec:Appendix Case}.

To accurately prompt LLMs for generating high-quality negatives, we propose a multi-attribute self-reflection strategy for prompt construction. This strategy includes five distinct types of definitions designed to impose rigorous constraints, along with a \textit{one-shot} example comprising a query $q$ and its corresponding positive document $d^+$. These elements are grouped into three main categories: 

\begin{itemize}[leftmargin=*]
\item \textbf{context Constrain:} Three definitions are included to constrain negative context generation: ``Task Definition'', ``Hard Negatives Definition'' and ``Format Definition''. In ``Task Definition'' and ``Hard Negatives Definition'', generation task is explicitly described and 
we refer the widely accepted definition of hard negatives by \citet{robinson2020contrastive}, characterizing them as \textit{``appearing to address the user query at first glance but subtly diverging in a manner that does not truly answer the query or fulfill the user's informational needs.''}. This definition serves as a guiding framework for the LLMs to produce valid negatives. In ``Format Definition'', we instruct the LLM to generate outputs in JSON format to facilitate further processing.

\item \textbf{Attributes Constrain:} Synthetic data generated by LLMs often suffers from repetitiveness \cite{divekar2024synthesizrr}. \citet{yu2024large} proposes incorporating multiple attributes in the prompt can improve diversity. To address this, we introduce an ``Attribute Definition'' that includes three attributes: \textit{\{domain name\}}, \textit{\{difficulty level\}}, and \textit{\{generation length\}}. These control the content sub-domain, complexity, and length of the generation. A set of predefined candidates for each attribute is used, with one randomly selected for each prompt, ensuring diverse topics and content, and enhancing the richness and reliability of the generated negatives.

\item \textbf{Reflection Constrain:} To further improve the quality of generated negatives, we integrate a self-reflection process in ``Reasoning Definition''. This involves prompting the LLM to perform a COT reasoning step~\cite{wei2022chain} based on the negative definitions and attribute specifications, aiming to balance accuracy and diversity. Specifically, we include explicit instructions in the prompt: \textit{``Write the inference process step by step ... including how to associate from the `user\_query' and `positive\_document' to derive the hard-negative documents.''} Thus, this process compels LLMs to perform a reasoning step when generating negatives, thereby improving their quality. Our experiment in Section~\ref{sec:Prompt design experiment} further validates this.

\end{itemize}

\subsection{Hybird Strategy for Negatives Mixing} \label{sec:Hybird Strategy}
In this section, we detail the process of incorporating synthetic negatives into the training workflow.

Prior studies treat synthetic samples as external data augmentations to enhance performance. For example, \citet{dai2022promptagator} generate queries to augment query-doc pairs, \citet{li2024synthetic} generate QA pairs to enrich datasets, and other works like \cite{wang2023improving, ma2023pre}, follow similar approaches. We refer to these methods as the ``Direct Mix Strategy'', as depicted in Figure~\ref{fig:mix strategy} (a). 

However, we find that training solely with LLM-generated negatives using this strategy leads to a decline in performance, since synthetic negatives tend to be more challenging, and during training, they would yield stronger optimization signals, which can lead to training instability. Previous study~\cite{zhou2022simans} indicates that larger gradient variance during training can destabilize parameter optimization and impede model convergence, and also validated by \citet{johnson2018training} and \citet{katharopoulos2018not}.

\begin{figure}[t]
    \centering
    \vspace{-2mm}
    \includegraphics[width=0.85\linewidth]{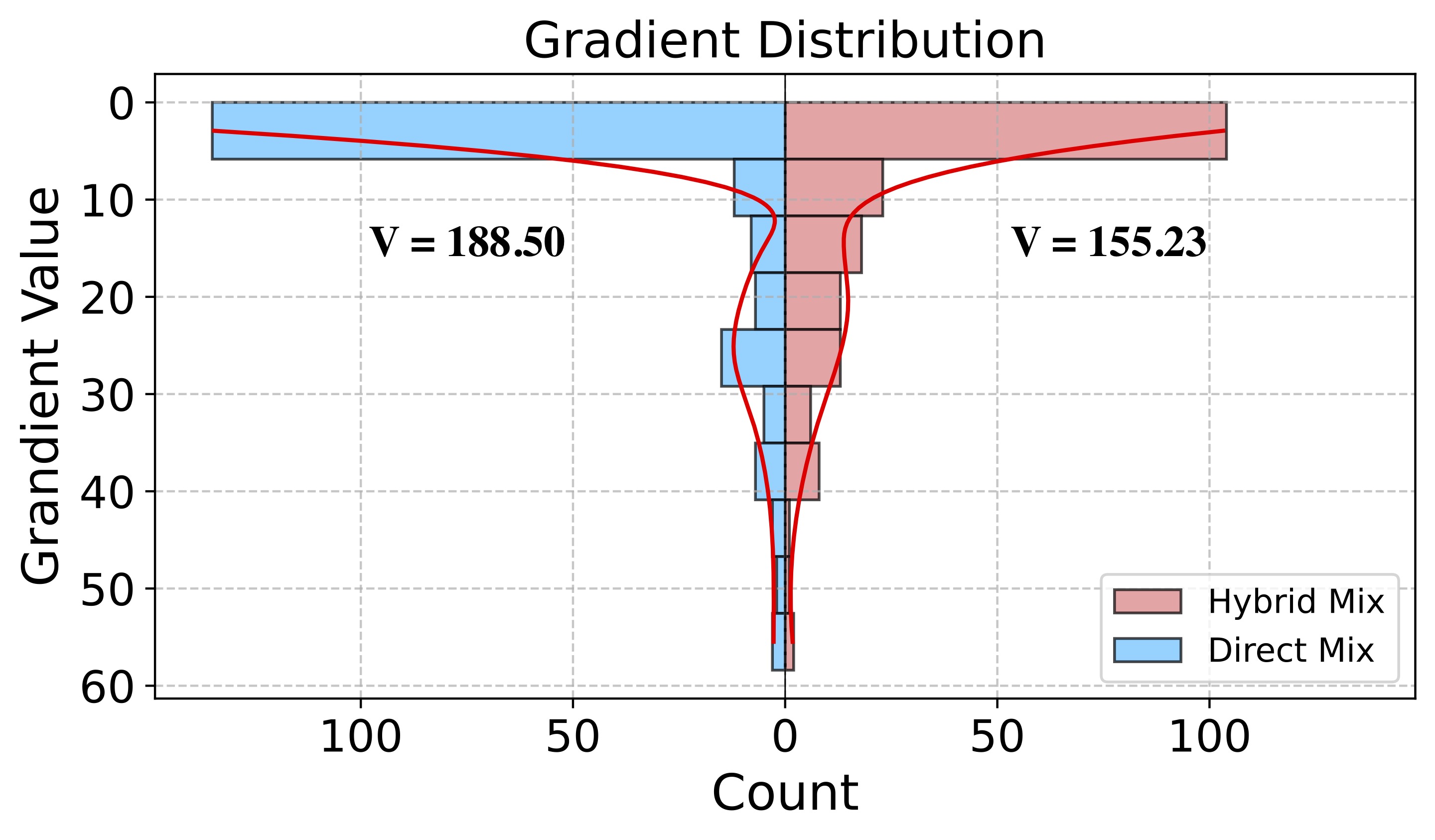}
    \vspace{-4mm}
    \caption{Gradient distribution in the first epoch in training for Scifact. Hybrid mix strategy owns a more even gradient distribution and lower variance.}
    \vspace{-4mm}
    \label{fig:gradient}
\end{figure}

Thus, to address this issue, we introduce a ``Hybrid Mix Strategy'', which blends generative negatives with conventional retrieved negatives at the instance level during training, as shown in Figure~\ref{fig:mix strategy} (b).
Specifically, for each query-positive pair, we generate one synthetic negative and use a negative retriever $g$ to retrieve the remaining negatives. Unlike the ``Direct Mixing Strategy'', our hybrid strategy bring a more stable gradient during training and improved performance. Gradients distribution are recorded in Figure~\ref{fig:gradient}, showing that compared with the ``Direct Mixing Strategy'', our hybrid strategy exhibits a more even distribution of gradient values and besides, the gradient variance is $\mathbb{V}=155.23$, much less than $\mathbb{V}=188.5$ for direct mixing, providing solid evidence that our approach stabilizes the training process and owns perspective to enhance performance.

\subsection{Model Fine-tuning}
In this section, we describe how we use hybrid samples to fine-tune the DR models.

For our experiment, we select different DR backbone models to continue fine-tuning with our synthetic dataset, which is derived from the previous phase. We use InfoNCE~\cite{oord2018representation} as the loss function for contrastive learning, defined as:
\begin{equation}
\min \mathcal{L} = - \log \frac{f(q, d^+)}{f(q, d^+) + \sum_{d^- \in \mathcal{D}_h^-} f(q, d^-)}
\end{equation}
here, $\mathcal{D}_h^-$ denotes the set of hybrid negatives, including both synthetic and retrieved negatives. Function $f$ is used to compute the relation between $q$ and $d$, formulated as:
\begin{equation}\label{equ:f function}
f (q, d) = \exp \left(\frac{1}{\tau} \cdot s(q,d) \right)
\end{equation}
in our work, we use the dot product to calculate similarity. Additionally, a temperature scaling factor, $\tau$, is introduced to dynamically control the discrepancy between the $q$ and $d$.
\section{Experiment}\label{sec:Experiment}
\subsection{Datasets and Experiment Settings}
We select five datasets from BEIR~\cite{thakur2021beir} for validation. We follow the default splitting setting of these datasets. Experimental details are depicted in Appendix~\ref{sec:Appendix Experiment Details}.

\subsection{Baselines}
We compare SyNeg with two baseline categories: 
\begin{itemize}[leftmargin=*]

\item \textbf{Retrieval Baselines:} To test the compatibility of Syneg with different DR models, we include three types of models: \textbf{1). Sparse model}: BM25; \textbf{2). Dense model}: COLBERT, ANCE, DPR, coCondenser, Contriever, Unifier, MoMA, COCO-DR, and GTR. This also includes text-embedding models such as BGE, E5, and GTE-Qwen2; \textbf{3). Dense model with synthetic data}: This type includes jina-embedding, mE5-large-instruct.
\item \textbf{Hard Negative Sampling Baselines:} We compare Syneg against several hard negative sampling baselines: \textbf{Random Neg}, sampling random documents from the corpus as negatives; \textbf{BM25 Neg}, using BM25 to retrieve top-ranked documents as negatives; \textbf{ANCE Neg}, employing a warm-up dense model to retrieve negatives; \textbf{BGE Neg}, using BGE to retrieve negatives; \textbf{SimANS}, which samples negatives similar to positives, focusing on ambiguous negatives; and \textbf{ADORE}, which uses dynamic negatives for dense retrieval.

\end{itemize}

\subsection{Main Results}

\begin{table*}[htbp]
  \centering
  \vspace{-5mm}
  \scalebox{0.8}{
    \begin{tabular}{lc|cccccl}
    \toprule
        \textbf{Model} & \textbf{Param.} & \textbf{FIQA}  & \textbf{SciFact} & \textbf{Quora} &\textbf{{HotpotQA}} & \textbf{{MS Marco}} &\textbf{{Avg.}}\\
    
    \midrule
    
    {\textit{Sparse Model}} & & & & &&  &\\
     BM25~\cite{gao2021complement}  & -     & 23.6 & 66.5 & 78.9 &60.3&  22.8 &50.4\\
    
    \midrule
    
    {\textit{Dense Model}} & & & & &&  &\\
     COLBert~\cite{khattab2020colbert} & 110M  & 31.7 & 67.1 & 85.4 &59.3 & 40.1 &56.7\\
     ANCE~\cite{xiong2020approximate}& 110M& 29.5 & 50.7 & 85.2 &45.6 & 38.8 &49.9\\
     DPR~\cite{karpukhin2020dense}   & 110M  & 27.4 & 47.5 & 84.2 &39.1 & 17.7 &43.1\\
     coCondenser~\cite{gao2021unsupervised}& 110M& 28.5 & 60.0 & 86.3 &56.6 & - &
-\\
     Contriever~\cite{izacard2021unsupervised} & 110M  & 32.9 & 67.7 &86.5 &63.8 & 40.7 &58.3\\
     Unifier~\cite{shen2023unifier} & 110M  & 31.1 & 68.6 &- &66.1 &- &
-\\
     MoMA (T5-ANCE)~\cite{ge2023augmenting}& 110M*2& 32.0 & 63.2 &84.7 &53.9 & - &
-\\
     MoMA (coCondenser)~\cite{ge2023augmenting}& 110M*2& 32.9 & 63.0 &84.3 &58.9 &- &
-\\
     COCO-DR~\cite{yu2022coco} & 110M  & 30.7 & 70.9 &86.7 &61.6 &41.9 &58.3\\
     GTR-Base~\cite{ni2021large}& 110M& 34.9 & 60.0 & 88.1 &53.5 &- &
-\\
     GTR-large~\cite{ni2021large}& 335M& 42.4 & 63.9 &89.0 &57.9 &- &
-\\
     E5-base~\cite{wang2024multilingual}& 278M  & 38.1& 69.3 & 87.6&68.5 & 42.2 &61.1\\
     BGE-large~\cite{bge_embedding}& 335M  & 45.0 & 72.4 & 88.6 &74.1&42.4 &64.5
\\
    \midrule
    {\textit{Model with Syntheic Data}} & & & & &&  &\\
    jina-v2~\cite{gunther2023jina}& 137M& 41.5 & 66.6 &88.2 &61.3 &40.9 &59.7
\\
    mE5-large-instruct~\cite{wang2024multilingual}& 560M& 47.7 & 71.8 &89.1 &69.3& 40.4 &63.6\\
    
    \midrule
    
    {\textit{Ours}} & & & & & &  &\\
     \textbf{COCO-DR + SyNeg}& 110M  & 33.8& 79.4& 87.1 &63.5& 42.2 & 61.2$^{\textbf{+2.9}}$\\
     
     \textbf{E5-base+ SyNeg}& 278M  & 38.6 & 77.0 & 88.5 &69.7& 42.4  & 63.2$^{\textbf{+2.1}}$\\
     
     \textbf{BGE-large+ SyNeg} & 335M  & 48.2 & 82.4 & 89.3 &75.0& 42.8 &67.5$^{\textbf{+3.0}}$\\

    \bottomrule
    
    \end{tabular}}
    \vspace{-2mm}
      \caption{Performance comparison of different dense retrieval models (<1B) in NDCG@10, with baseline results sourced from the original papers or the MTEB leaderboard~\cite{muennighoff2022mteb}.}
      \vspace{-2mm}
      \label{tab:model performance <1B}
      \end{table*}

\begin{table}[htbp]
  \centering
  \vspace{-2mm}
    \scalebox{0.7}{
    \begin{tabular}{lc|cc}
    \toprule
    \textbf{Model} & \textbf{Param.} & \textbf{Quora}  &\textbf{HotpotQA}\\
    \midrule
    
    {\textit{Dense Model}} & & & \\
    GTE-Qwen2-1.5B~\cite{li2023towards}& 1.7B  & 90.1  &   68.9\\
    GTE-Qwen2-7B~\cite{li2023towards}& 7.6B  & 89.6   &  73.0\\
    \midrule
    
    {\textit{Ours}} & & \\
   \textbf{GTE-Qwen2-1.5B + SyNeg}& 1.7B  & 90.3   & 69.3 \\
   \textbf{GTE-Qwen2-7B + SyNeg}& 7.6B  & 89.9  & 74.1 \\

    \bottomrule
    \end{tabular}
    }
    \vspace{-3mm}
    \caption{ {Performance comparison among different dense retrieval models (>1B) in NDCG@10.}}
    \vspace{-4mm}

  \label{tab:model performance >1B}
\end{table}

To examine whether our proposed pipeline consistently enhances performance across different DR models on different datasets. We select three small models(<1B) and two large models (>1B), results are shown in Table~\ref{tab:model performance <1B} and Table~\ref{tab:model performance >1B}.

\begin{itemize}[leftmargin=*]
  \item Results in Table~\ref{tab:model performance <1B} reveal that SyNeg significantly improves the performance of DR models across different datasets, with average increases of +2.9/+2.1/+3.0 points for COCO-DR, E5, and BGE, respectively. This improvement is attributed to the combined effect of introducing synthetic negatives during downstream training and employing a hybrid strategy to stabilize the training process. Notably, for knowledge-intensive datasets (e.g., Scifact), the improvement is particularly significant. Additionally, our method also yields substantial  gains on much larger datasets (e.g., HotpotQA, MS Marco).
  
  \item {In Table~\ref{tab:model performance >1B}, We validate our results on LLMs (>1B) using two large datasets, Quora and HotpotQA, with two versions of GTE-Qwen2 (1.5B and 7B). Consistently, our methods significantly enhance LLM performance, highlighting the superiority of our method design.}

\end{itemize}

\subsection{Sampling Strategies Comparison}

\begin{table*}[htbp]
  \centering
  \vspace{-4mm}
  \scalebox{0.52}{
    \begin{tabular}{l|ccc|ccc|ccc|ccc}
    \toprule
    \multirow{2}{*}{\textbf{Model}} & \multicolumn{3}{c|}{\textbf{FIQA}} & \multicolumn{3}{c|}{\textbf{SciFact}} & \multicolumn{3}{c|}{\textbf{Quora}} & \multicolumn{3}{c}{\textbf{{HotpotQA}}} \\
\cmidrule{2-13}          & \textbf{NDCG@10} & \textbf{R@10} & \textbf{MRR@10} & \textbf{NDCG@10} & \textbf{R@10} & \textbf{MRR@10} & \textbf{NDCG@10} & \textbf{R@3} & \textbf{MRR@10} & \textbf{NDCG@10} & \textbf{R@10} & \textbf{MRR@10}\\
    \midrule
    Random Neg~\cite{huang2020embedding} & 39.5  & 46.3  & 43.7  & 79.8  & 91.8  & 76.3  & 86.7  & 86.3  & 85.3 & 69.9 & 72.7 & 84.1 
\\
    BM25 Neg~\cite{gao2021complement} & 40.2  & 48.1  & 47.9  & 80.5  & 91.6  & 77.3  & 88.2  & 87.3  & 87.5 & 69.7 & 72.1 & 83.8 
\\
    ANCE Neg~\cite{xiong2020approximate} & 41.1  & 47.6  & 49.9  & 80.4  & 91.0  & 77.5  & 88.6  & 87.6  & 87.9 & 70.5 & 73.2 & 85.2 
\\
    BGE Neg~\cite{bge_embedding} & \underline{42.8}  & \underline{50.0}  & \underline{51.3}  & 81.1  & 92.1  & 78.2  & 88.8  & 87.6  & 88.0 & \underline{70.9} & \underline{75.4} & \underline{85.6} 
\\
    SimANS + BGE Neg~\cite{zhou2022simans} & 35.9  & 41.7  & 43.8  & \underline{82.2}  & \underline{92.4}  & 79.0  & \underline{89.1}  & 87.9  & \underline{88.5} & 70.0 & 73.0 & 83.9 
\\
    ADORE + BGE Neg~\cite{zhan2021optimizing} & 34.7  & 40.9  & 42.3  & 81.9  & 91.5  & \underline{79.1}  & 89.0  & \underline{88.0}  & 88.3 & 69.5 & 73.4 & 82.4 
\\
    \midrule
    \textbf{SyNeg (Ours)} & \textbf{48.2} & \textbf{53.5} & \textbf{53.1} & \textbf{82.4} & \textbf{93.2} & \textbf{79.5} & \textbf{89.3} & \textbf{88.3} & \textbf{88.7} &  \textbf{75.0} &  \textbf{80.1} &  \textbf{87.7}
\\
    \bottomrule
    \end{tabular}
    }
    \vspace{-3mm}
    \caption{Comparison of sampling strategy results. Best results are shown in \textbf{bold}, and second-best results are \underline{underlined}. }
    \vspace{-3mm}
  \label{tab:sample perofrmance}
\end{table*}

We compare several negative sampling strategies, including Random, BM25, ANCE, BGE-based retriever,  SimANS, and ADORE. The implementation settings of last two methods are provided in Appendix~\ref{sec:Appendix Experiment Details}. The overall results are shown in Table~\ref{tab:sample perofrmance}.
\begin{itemize}[leftmargin=*]
  \item Syneg outperforms various sampling methods in negative retrieval tasks. Random Negatives sampling, due to its wide sampling range, often selects simple negatives, leading to suboptimal results. The BM25 model, using sparse retrieval, offers slight improvements over Random Negatives but remains inadequate. ANCE, with its warm-up training phase, surpasses both Random Negatives and BM25. BGE, the current sota for negative retrieval, excels, particularly on the FIQA and HotpotQA. SimANS, which employs ambiguous negatives, achieves notable success on the SciFact and Quora, while ADORE, using a dynamic negative strategy, also delivers strong results. Our approach, combining synthetic and hybrid negative strategy, produces higher-quality negatives for model fine-tuning, leading to superior performance across baselines. We also observe that the improvements vary by dataset, likely due to the distribution differences. Our method, leveraging LLMs to generate hard negatives and enhance retrieval performance, is particularly effective on FIQA, requiring a higher degree of discriminative ability due to their specialized knowledge domains. 

\end{itemize}

\subsection{Analysis} \label{sec:ablation Experiment} 
In this section, we present four analytical parts to provide a deeper exploration of our proposed method.

\subsubsection{Are the generative negatives much harder?} \label{sec:harder experiment}

\begin{figure}[t]
\centering
\vspace{-2mm}
{\subfigure{\includegraphics[width=0.8\linewidth]{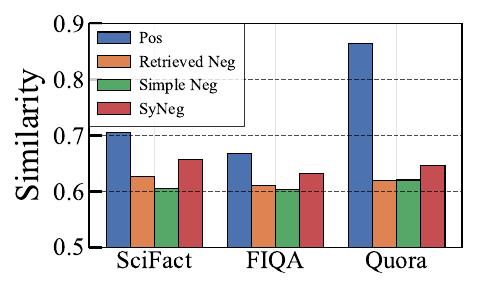}}}
\vspace{-2mm}
\caption{Similarity test results.}
\vspace{-1mm}
\label{fig:similarity}
\end{figure}

To quantitatively evaluate the difficulty of generative negatives, we conduct similarity tests to intuitively assess the quality of the generated negatives, as shown in Figure~\ref{fig:similarity}. Specifically, we calculate the average similarity scores between queries and four categories of documents: positive documents, BGE-retrieved documents, simple negatives (generated using only a basic negative definition in the prompt, see Appendix~\ref{sec:Simple Prompt}), and negatives generated using our proposed method. Our results reveal that generative negatives produced by our method are significantly more similar to positives than other types of negatives, indicating that they are substantially harder. Additionally, we provide case examples in Appendix~\ref{sec:Appendix Case} to illustrate specific instances for each dataset.

\subsubsection{{Analysis of Negative Generation}} \label{sec:Prompt design experiment}

\begin{table}[t]
  \centering
  \vspace{-1mm}
    \scalebox{0.8}{
    \begin{tabular}{lccc}
    \toprule
         \textbf{Methods/NDCG@10} & \textbf{SciFact} & \textbf{FIQA} & \textbf{Quora}  \\
    \midrule
    \textit{w/o} Attributes+Reflection  & 76.5  & 40.2  & 84.3   \\
    \textit{w/o} Attributes & 80.0  & 44.7  & 88.9  \\
    \midrule
    \textbf{SyNeg (Ours)} & \textbf{82.4} & \textbf{48.2} & \textbf{89.2} \\
    \bottomrule
    \end{tabular}
    }
    \vspace{-2mm}
    \caption{{Ablation  results for negative generation.}}
    \vspace{-4mm}
\label{tab:ablation generation}
\end{table}

{In Section~\ref{sec:hard-neg generation}, we propose a strategy that employs a multi-attribute self-reflection prompt to generate negatives via LLMs. To verify its effectiveness, we conduct ablation experiments to assess the impact of each component on the generation. Specifically, we ablate attributes and reflections components within experiment, regenerate negatives, and fine-tune them on the same retriever. The results are shown in Table~\ref{tab:ablation generation}.}

{Results show that both multi-attribute and self-reflection play important roles in negative generation. One key observation is that self-reflection is slightly more crucial for generating negatives in Scifact and Quora, indicating that explicit reasoning during generation with LLMs is essential.}

\subsubsection{Analysis of Hybrid mixing Stretegy}

\begin{table}[t]
  \centering
  \vspace{-2mm}
  \scalebox{0.7}{
    \begin{tabular}{lccc}
    \toprule
   \textbf{Methods/NDCG@10}   & \textbf{SciFact} & \textbf{FIQA}  & \textbf{Quora} \\
    \midrule
    \textit{w/o FT} & 72.4  & 45.0  & 88.6  \\

    \textit{w/ FT} - Pure Synthetic Neg & 77.6  & 41.3  & 85.0  \\

    \textit{w/ FT} - Direct Mixing & 81.1  & 41.7  & 87.3  \\
    \midrule
    \textit{w/ FT} - \textbf{Hybird mix strategy (Ours)} & \textbf{82.4}  & \textbf{48.2}  & \textbf{89.3}  \\
    \bottomrule
    \end{tabular}}
    \vspace{-2mm}
    \caption{Different negatives mixing strategies comparison.``FT'' denotes Fine-Tuning.}
    \vspace{-2mm}
  \label{tab:mixstrategy}
\end{table}

In Section~\ref{sec:Hybird Strategy}, we show that a direct mixing strategy leads to unstable training and decreased performance. In this part, we conduct extensive experiments to compare different negative mixing strategies and their impact on overall performance, as shown in Table~\ref{tab:mixstrategy}. The tests include ``w/o FT'', representing the zero-shot experiment with the original BGE retrieval performance, and two comparison experiments: fine-tuning with purely LLM-generated negatives and using the direct mixing strategy for fine-tuning. There are two key findings: First, we find that fine-tuning with purely synthetic negatives is not effective and can even harm performance. We hypothesize this is because synthetic negatives are too challenging for the model to learn from effectively. Second, direct mixing does not consistently improve performance across tasks, as it causes unstable training gradients between different batches. Our proposed mixing strategy addresses these issues and achieves superior performance.

\subsubsection{Analysis of Parameters}
This section examines the impact of two hyperparameters: the temperature factor $\tau$ during optimization and the ratio of synthetic negatives in the training samples.

\vspace{1mm}
\textbf{\textit{{(1). Temperature Experiment}}}
\vspace{1mm}

\begin{figure}[t]
	\centering
    \vspace{-5mm}
    {\subfigure{\includegraphics[width=0.49\linewidth]{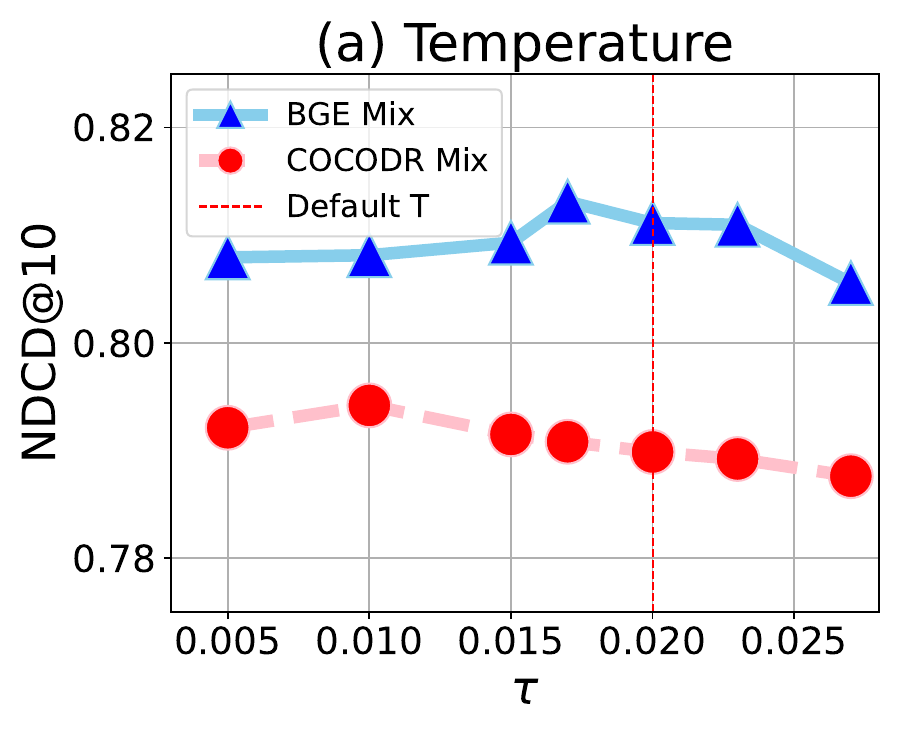}}}
	{\subfigure{\includegraphics[width=0.49\linewidth]{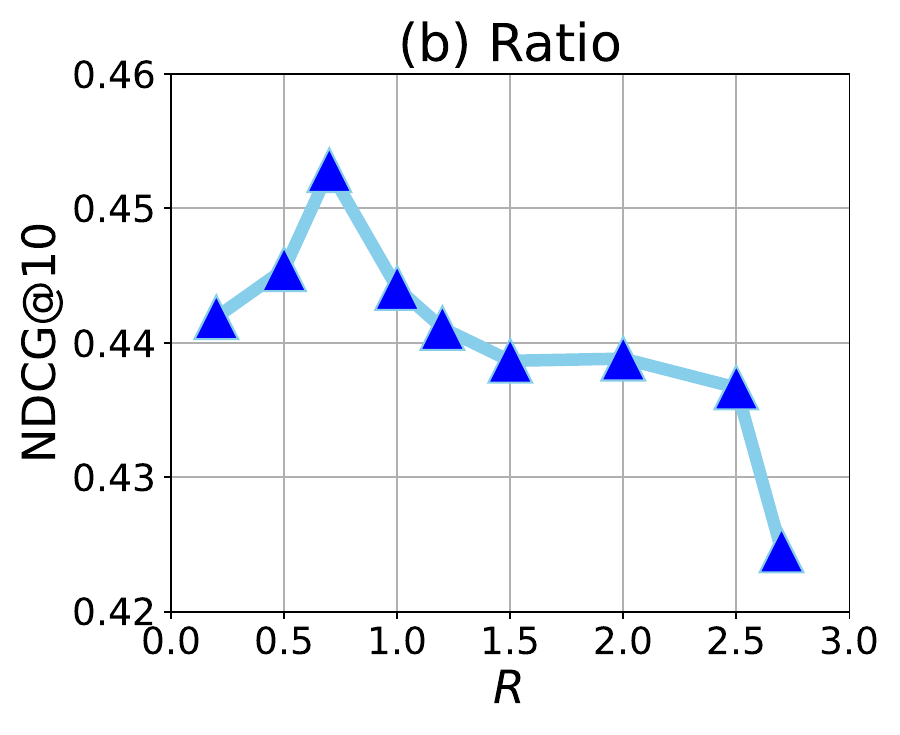}}}
    \vspace{-8mm}
	\caption{Hyper-parameter experiment results.}
	\vspace{-5mm}
 \label{fig:hyper}
\end{figure}

As shown in Equation~\eqref{equ:f function}, the temperature factor $\tau$ balances the contributions of positive and negative sample pairs to the loss. Specifically, a lower $\tau$ increases the contrast between positive and negative pairs, enhancing the model's ability to distinguish between them. In the experiment by \citet{wang2023improving}, $\tau$ is set to 0.02. In our setting, where generative negatives are significantly harder, controlling $\tau$ is crucial during training. Experiment results are shown in Figure~\ref{fig:hyper} (a). We conduct two experiments with BGE and COCODR on SciFact, respectively. We find that for synthetic negatives, a lower $\tau$ yields optimal performance, as expected. For example, for BGE, $\tau = 0.017$ achieves the best performance, while for COCODR, $\tau = 0.01$ is optimal. Based on these results, we select a lower $\tau$ to help the model effectively distinguish between positives and negatives.

\vspace{1mm}
\textbf{\textit{{(2). Ratio of Synthetic Negatives Mixing Experiment}}}
\vspace{1mm}

In this part, we examine the optimal amount of synthetic negatives required during our negative mixing process. Specifically, we aim to determine the ratio of synthetic negatives that yields the best results. We define the ratio factor $R$ as the ratio of synthetic negative documents to positive documents in the entire dataset, mathematically expressed as $R = \frac{num(\mathcal{D}^-_{syn})}{num(\mathcal{D}^+)}$. The results, shown in Figure~\ref{fig:hyper} (b), indicate that a small ratio of synthetic negatives (top at 0.7) significantly enhances performance. However, as $R$ increases, the dataset becomes dominated by synthetic negatives, leading to a performance drop. This suggests that, given the high cost of generating negatives with LLMs, a carefully selected small proportion of synthetic negatives is promising for both cost-effective and yields optimal performance.

\section{Related Work}
\subsection{Synthetic Data Generation}

Synthetic data generation gains much attention recently especially in IR. We categorize works into three types: 1) Query Generation:  GPL~\cite{wang2021gpl} generates queries and pseudo labels using a cross-encoder. InPars~\cite{bonifacio2022inpars} and Promptagator~\cite{dai2022promptagator} apply few-shot strategies with LLMs to create queries for unlabeled documents, which are then used for model training. 2) Document Generation: Using LLMs to create synthetic docs, like Query2Doc~\cite{wang2023query2doc}, generating pseudo-documents via few-shot prompting for query expansion. 3) Instance Generation: Simultaneously generates queries and documents, like~\cite{wang2023improving}, creating query-doc pairs for fine-tuning. Our method falls into the second category, generating hard negatives to enhance fine-tuning.negatives to enhance fine-tuning.

\subsection{Hard-negative Sampling}

Hard negative sampling is critical for enhancing DR performance, we categorize into five types: 1). Random Sampling: The most common strategy, using random documents or other query-relevant documents as negatives (in-batch negatives), as seen in~\cite{huang2020embedding, ding2020rocketqa}.
2). Sparse Retrieval: This strategy leverages sparse retrievers like BM25 to fetch hard negatives, such as in~\cite{gao2021complement, karpukhin2020dense}, where BM25 retrieves top-ranked candidates as negatives.
3). Warm-Up Training Models: Approaches like ANCE~\cite{xiong2020approximate} and REALM~\cite{guu2020retrieval} use pre-trained dense retrievers for negative sampling. ANCE-Tele~\cite{sun2022reduce} further improves ANCE by combining three types of negatives.
4). Distribution Optimization: Negatives are chosen based on empirical or manually defined distributions. SimANS~\cite{zhou2022simans} samples ambiguous negatives near the positive document, while Trisampler~\cite{yang2024trisampler} selects informative negatives based on the "quasi-triangular principle."
5). Dynamic Negatives: Proposed by~\citet{zhan2021optimizing}, this method refreshes negatives dynamically, selecting different ones for each query during training.
Distinct from these methods, our approach utilizes LLMs to generate hard negatives, integrating them into the training process to further improve model performance.
\section{Conclusion}
We introduce SyNeg, a novel approach for generating and using synthetic hard negatives in dense retrieval with LLMs. Firstly, we propose a multi-attribute self-reflective strategy for generating high-quality hard negatives, secondly, a hybrid strategy that integrates LLM-generated negatives with retrieved negatives for stabilizing training. Extensive evaluations showing SyNeg's superiority. Future work could explore using open-source LLMs, enhancing negative diversity, and applying SyNeg to more Challenging IR tasks.
\section{Limitations}
In this work, we propose Syneg, an innovative pipeline that includes synthetic negative generation and training. Despite its promising performance on the BEIR task, one limitation is our reliance on API-based LLMs. Further exploration of open-source LLMs for synthetic generation, although it may result in a decrease in performance, is crucial to consider due to budgetary constraints. Additionally, our current setup focuses solely on text retrieval. The potential to generate samples in multimodal settings (e.g., text-to-image) also warrants exploration to further validate the effectiveness of our approach.

\bibliography{custom}

\begin{thebibliography}{44}
\providecommand{\natexlab}[1]{#1}

\bibitem[{Almeida and Matos(2024)}]{almeida2024exploring}
Tiago Almeida and S{\'e}rgio Matos. 2024.
\newblock Exploring efficient zero-shot synthetic dataset generation for information retrieval.
\newblock In \emph{Findings of the Association for Computational Linguistics: EACL 2024}, pages 1214--1231.

\bibitem[{Bonifacio et~al.(2022)Bonifacio, Abonizio, Fadaee, and Nogueira}]{bonifacio2022inpars}
Luiz Bonifacio, Hugo Abonizio, Marzieh Fadaee, and Rodrigo Nogueira. 2022.
\newblock Inpars: Data augmentation for information retrieval using large language models.
\newblock \emph{arXiv preprint arXiv:2202.05144}.

\bibitem[{Chen et~al.(2024)Chen, Xiao, Zhang, Luo, Lian, and Liu}]{chen2024bge}
Jianlv Chen, Shitao Xiao, Peitian Zhang, Kun Luo, Defu Lian, and Zheng Liu. 2024.
\newblock Bge m3-embedding: Multi-lingual, multi-functionality, multi-granularity text embeddings through self-knowledge distillation.
\newblock \emph{arXiv preprint arXiv:2402.03216}.

\bibitem[{Craswell et~al.(2020)Craswell, Mitra, Yilmaz, Campos, and Voorhees}]{craswell2020overview}
Nick Craswell, Bhaskar Mitra, Emine Yilmaz, Daniel Campos, and Ellen~M Voorhees. 2020.
\newblock Overview of the trec 2019 deep learning track.
\newblock \emph{arXiv preprint arXiv:2003.07820}.

\bibitem[{Dai et~al.(2022)Dai, Zhao, Ma, Luan, Ni, Lu, Bakalov, Guu, Hall, and Chang}]{dai2022promptagator}
Zhuyun Dai, Vincent~Y Zhao, Ji~Ma, Yi~Luan, Jianmo Ni, Jing Lu, Anton Bakalov, Kelvin Guu, Keith~B Hall, and Ming-Wei Chang. 2022.
\newblock Promptagator: Few-shot dense retrieval from 8 examples.
\newblock \emph{arXiv preprint arXiv:2209.11755}.

\bibitem[{Ding et~al.(2020)Ding, Liu, Liu, Ren, Zhao, Dong, Wu, and Wang}]{ding2020rocketqa}
Yingqi Qu~Yuchen Ding, Jing Liu, Kai Liu, Ruiyang Ren, Xin Zhao, Daxiang Dong, Hua Wu, and Haifeng Wang. 2020.
\newblock Rocketqa: An optimized training approach to dense passage retrieval for open-domain question answering.
\newblock \emph{arXiv preprint arXiv:2010.08191}.

\bibitem[{Divekar and Durrett(2024)}]{divekar2024synthesizrr}
Abhishek Divekar and Greg Durrett. 2024.
\newblock Synthesizrr: Generating diverse datasets with retrieval augmentation.
\newblock \emph{arXiv preprint arXiv:2405.10040}.

\bibitem[{Gao and Callan(2021)}]{gao2021unsupervised}
Luyu Gao and Jamie Callan. 2021.
\newblock Unsupervised corpus aware language model pre-training for dense passage retrieval.
\newblock \emph{arXiv preprint arXiv:2108.05540}.

\bibitem[{Gao et~al.(2021)Gao, Dai, Chen, Fan, Van~Durme, and Callan}]{gao2021complement}
Luyu Gao, Zhuyun Dai, Tongfei Chen, Zhen Fan, Benjamin Van~Durme, and Jamie Callan. 2021.
\newblock Complement lexical retrieval model with semantic residual embeddings.
\newblock In \emph{Advances in Information Retrieval: 43rd European Conference on IR Research, ECIR 2021, Virtual Event, March 28--April 1, 2021, Proceedings, Part I 43}, pages 146--160. Springer.

\bibitem[{Gao et~al.(2023)Gao, Xiong, Gao, Jia, Pan, Bi, Dai, Sun, and Wang}]{gao2023retrieval}
Yunfan Gao, Yun Xiong, Xinyu Gao, Kangxiang Jia, Jinliu Pan, Yuxi Bi, Yi~Dai, Jiawei Sun, and Haofen Wang. 2023.
\newblock Retrieval-augmented generation for large language models: A survey.
\newblock \emph{arXiv preprint arXiv:2312.10997}.

\bibitem[{Ge et~al.(2023)Ge, Xiong, Rosset, Overwijk, Han, and Bennett}]{ge2023augmenting}
Suyu Ge, Chenyan Xiong, Corby Rosset, Arnold Overwijk, Jiawei Han, and Paul Bennett. 2023.
\newblock Augmenting zero-shot dense retrievers with plug-in mixture-of-memories.
\newblock \emph{arXiv preprint arXiv:2302.03754}.

\bibitem[{G{\"u}nther et~al.(2023)G{\"u}nther, Ong, Mohr, Abdessalem, Abel, Akram, Guzman, Mastrapas, Sturua, Wang et~al.}]{gunther2023jina}
Michael G{\"u}nther, Jackmin Ong, Isabelle Mohr, Alaeddine Abdessalem, Tanguy Abel, Mohammad~Kalim Akram, Susana Guzman, Georgios Mastrapas, Saba Sturua, Bo~Wang, et~al. 2023.
\newblock Jina embeddings 2: 8192-token general-purpose text embeddings for long documents.
\newblock \emph{arXiv preprint arXiv:2310.19923}.

\bibitem[{Guu et~al.(2020)Guu, Lee, Tung, Pasupat, and Chang}]{guu2020retrieval}
Kelvin Guu, Kenton Lee, Zora Tung, Panupong Pasupat, and Mingwei Chang. 2020.
\newblock Retrieval augmented language model pre-training.
\newblock In \emph{International conference on machine learning}, pages 3929--3938. PMLR.

\bibitem[{Huang et~al.(2020)Huang, Sharma, Sun, Xia, Zhang, Pronin, Padmanabhan, Ottaviano, and Yang}]{huang2020embedding}
Jui-Ting Huang, Ashish Sharma, Shuying Sun, Li~Xia, David Zhang, Philip Pronin, Janani Padmanabhan, Giuseppe Ottaviano, and Linjun Yang. 2020.
\newblock Embedding-based retrieval in facebook search.
\newblock In \emph{Proceedings of the 26th ACM SIGKDD International Conference on Knowledge Discovery \& Data Mining}, pages 2553--2561.

\bibitem[{Izacard et~al.(2021)Izacard, Caron, Hosseini, Riedel, Bojanowski, Joulin, and Grave}]{izacard2021unsupervised}
Gautier Izacard, Mathilde Caron, Lucas Hosseini, Sebastian Riedel, Piotr Bojanowski, Armand Joulin, and Edouard Grave. 2021.
\newblock Unsupervised dense information retrieval with contrastive learning.
\newblock \emph{arXiv preprint arXiv:2112.09118}.

\bibitem[{Johnson and Guestrin(2018)}]{johnson2018training}
Tyler~B Johnson and Carlos Guestrin. 2018.
\newblock Training deep models faster with robust, approximate importance sampling.
\newblock \emph{Advances in Neural Information Processing Systems}, 31.

\bibitem[{Karpukhin et~al.(2020)Karpukhin, O{\u{g}}uz, Min, Lewis, Wu, Edunov, Chen, and Yih}]{karpukhin2020dense}
Vladimir Karpukhin, Barlas O{\u{g}}uz, Sewon Min, Patrick Lewis, Ledell Wu, Sergey Edunov, Danqi Chen, and Wen-tau Yih. 2020.
\newblock Dense passage retrieval for open-domain question answering.
\newblock \emph{arXiv preprint arXiv:2004.04906}.

\bibitem[{Katharopoulos and Fleuret(2018)}]{katharopoulos2018not}
Angelos Katharopoulos and Fran{\c{c}}ois Fleuret. 2018.
\newblock Not all samples are created equal: Deep learning with importance sampling.
\newblock In \emph{International conference on machine learning}, pages 2525--2534. PMLR.

\bibitem[{Khattab and Zaharia(2020)}]{khattab2020colbert}
Omar Khattab and Matei Zaharia. 2020.
\newblock Colbert: Efficient and effective passage search via contextualized late interaction over bert.
\newblock In \emph{Proceedings of the 43rd International ACM SIGIR conference on research and development in Information Retrieval}, pages 39--48.

\bibitem[{Lee et~al.(2024)Lee, Dai, Ren, Chen, Cer, Cole, Hui, Boratko, Kapadia, Ding et~al.}]{lee2024gecko}
Jinhyuk Lee, Zhuyun Dai, Xiaoqi Ren, Blair Chen, Daniel Cer, Jeremy~R Cole, Kai Hui, Michael Boratko, Rajvi Kapadia, Wen Ding, et~al. 2024.
\newblock Gecko: Versatile text embeddings distilled from large language models.
\newblock \emph{arXiv preprint arXiv:2403.20327}.

\bibitem[{Li et~al.(2024)Li, Dong, Tang, Wang, Zhang, Huang, Huang, Huang, Huang, Zhang et~al.}]{li2024synthetic}
Haoran Li, Qingxiu Dong, Zhengyang Tang, Chaojun Wang, Xingxing Zhang, Haoyang Huang, Shaohan Huang, Xiaolong Huang, Zeqiang Huang, Dongdong Zhang, et~al. 2024.
\newblock Synthetic data (almost) from scratch: Generalized instruction tuning for language models.
\newblock \emph{arXiv preprint arXiv:2402.13064}.

\bibitem[{Li et~al.(2023)Li, Zhang, Zhang, Long, Xie, and Zhang}]{li2023towards}
Zehan Li, Xin Zhang, Yanzhao Zhang, Dingkun Long, Pengjun Xie, and Meishan Zhang. 2023.
\newblock Towards general text embeddings with multi-stage contrastive learning.
\newblock \emph{arXiv preprint arXiv:2308.03281}.

\bibitem[{Ma et~al.(2023)Ma, Wu, Wang, Lin, and Hu}]{ma2023pre}
Guangyuan Ma, Xing Wu, Peng Wang, Zijia Lin, and Songlin Hu. 2023.
\newblock Pre-training with large language model-based document expansion for dense passage retrieval.
\newblock \emph{arXiv preprint arXiv:2308.08285}.

\bibitem[{Muennighoff et~al.(2022)Muennighoff, Tazi, Magne, and Reimers}]{muennighoff2022mteb}
Niklas Muennighoff, Nouamane Tazi, Lo{\"\i}c Magne, and Nils Reimers. 2022.
\newblock \href {https://doi.org/10.48550/ARXIV.2210.07316} {Mteb: Massive text embedding benchmark}.
\newblock \emph{arXiv preprint arXiv:2210.07316}.

\bibitem[{Ni et~al.(2021)Ni, Qu, Lu, Dai, {\'A}brego, Ma, Zhao, Luan, Hall, Chang et~al.}]{ni2021large}
Jianmo Ni, Chen Qu, Jing Lu, Zhuyun Dai, Gustavo~Hern{\'a}ndez {\'A}brego, Ji~Ma, Vincent~Y Zhao, Yi~Luan, Keith~B Hall, Ming-Wei Chang, et~al. 2021.
\newblock Large dual encoders are generalizable retrievers.
\newblock \emph{arXiv preprint arXiv:2112.07899}.

\bibitem[{Oord et~al.(2018)Oord, Li, and Vinyals}]{oord2018representation}
Aaron van~den Oord, Yazhe Li, and Oriol Vinyals. 2018.
\newblock Representation learning with contrastive predictive coding.
\newblock \emph{arXiv preprint arXiv:1807.03748}.

\bibitem[{Reimers and Gurevych(2019)}]{reimers2019sentence}
Nils Reimers and Iryna Gurevych. 2019.
\newblock Sentence-bert: Sentence embeddings using siamese bert-networks.
\newblock \emph{arXiv preprint arXiv:1908.10084}.

\bibitem[{Robinson et~al.(2020)Robinson, Chuang, Sra, and Jegelka}]{robinson2020contrastive}
Joshua Robinson, Ching-Yao Chuang, Suvrit Sra, and Stefanie Jegelka. 2020.
\newblock Contrastive learning with hard negative samples.
\newblock \emph{arXiv preprint arXiv:2010.04592}.

\bibitem[{Shen et~al.(2023)Shen, Geng, Tao, Xu, Long, Zhang, and Jiang}]{shen2023unifier}
Tao Shen, Xiubo Geng, Chongyang Tao, Can Xu, Guodong Long, Kai Zhang, and Daxin Jiang. 2023.
\newblock Unifier: A unified retriever for large-scale retrieval.
\newblock In \emph{Proceedings of the 29th ACM SIGKDD Conference on Knowledge Discovery and Data Mining}, pages 4787--4799.

\bibitem[{Sun et~al.(2022)Sun, Xiong, Yu, Overwijk, Liu, and Bao}]{sun2022reduce}
Si~Sun, Chenyan Xiong, Yue Yu, Arnold Overwijk, Zhiyuan Liu, and Jie Bao. 2022.
\newblock Reduce catastrophic forgetting of dense retrieval training with teleportation negatives.
\newblock \emph{arXiv preprint arXiv:2210.17167}.

\bibitem[{Thakur et~al.(2021)Thakur, Reimers, R{\"u}ckl{\'e}, Srivastava, and Gurevych}]{thakur2021beir}
Nandan Thakur, Nils Reimers, Andreas R{\"u}ckl{\'e}, Abhishek Srivastava, and Iryna Gurevych. 2021.
\newblock Beir: A heterogenous benchmark for zero-shot evaluation of information retrieval models.
\newblock \emph{arXiv preprint arXiv:2104.08663}.

\bibitem[{Wang et~al.(2021)Wang, Thakur, Reimers, and Gurevych}]{wang2021gpl}
Kexin Wang, Nandan Thakur, Nils Reimers, and Iryna Gurevych. 2021.
\newblock Gpl: Generative pseudo labeling for unsupervised domain adaptation of dense retrieval.
\newblock \emph{arXiv preprint arXiv:2112.07577}.

\bibitem[{Wang et~al.(2024{\natexlab{a}})Wang, Yang, Huang, Yang, Majumder, and Wei}]{wang2023improving}
Liang Wang, Nan Yang, Xiaolong Huang, Linjun Yang, Rangan Majumder, and Furu Wei. 2024{\natexlab{a}}.
\newblock \href {https://doi.org/10.18653/v1/2024.acl-long.642} {Improving text embeddings with large language models}.
\newblock In \emph{Proceedings of the 62nd Annual Meeting of the Association for Computational Linguistics (Volume 1: Long Papers)}, pages 11897--11916, Bangkok, Thailand. Association for Computational Linguistics.

\bibitem[{Wang et~al.(2024{\natexlab{b}})Wang, Yang, Huang, Yang, Majumder, and Wei}]{wang2024multilingual}
Liang Wang, Nan Yang, Xiaolong Huang, Linjun Yang, Rangan Majumder, and Furu Wei. 2024{\natexlab{b}}.
\newblock Multilingual e5 text embeddings: A technical report.
\newblock \emph{arXiv preprint arXiv:2402.05672}.

\bibitem[{Wang et~al.(2023)Wang, Yang, and Wei}]{wang2023query2doc}
Liang Wang, Nan Yang, and Furu Wei. 2023.
\newblock Query2doc: Query expansion with large language models.
\newblock \emph{arXiv preprint arXiv:2303.07678}.

\bibitem[{Wei et~al.(2022)Wei, Wang, Schuurmans, Bosma, Xia, Chi, Le, Zhou et~al.}]{wei2022chain}
Jason Wei, Xuezhi Wang, Dale Schuurmans, Maarten Bosma, Fei Xia, Ed~Chi, Quoc~V Le, Denny Zhou, et~al. 2022.
\newblock Chain-of-thought prompting elicits reasoning in large language models.
\newblock \emph{Advances in neural information processing systems}, 35:24824--24837.

\bibitem[{Xiao et~al.(2023)Xiao, Liu, Zhang, and Muennighoff}]{bge_embedding}
Shitao Xiao, Zheng Liu, Peitian Zhang, and Niklas Muennighoff. 2023.
\newblock \href {https://arxiv.org/abs/2309.07597} {C-pack: Packaged resources to advance general chinese embedding}.
\newblock \emph{Preprint}, arXiv:2309.07597.

\bibitem[{Xiong et~al.(2020)Xiong, Xiong, Li, Tang, Liu, Bennett, Ahmed, and Overwijk}]{xiong2020approximate}
Lee Xiong, Chenyan Xiong, Ye~Li, Kwok-Fung Tang, Jialin Liu, Paul Bennett, Junaid Ahmed, and Arnold Overwijk. 2020.
\newblock Approximate nearest neighbor negative contrastive learning for dense text retrieval.
\newblock \emph{arXiv preprint arXiv:2007.00808}.

\bibitem[{Yang et~al.(2024)Yang, Shao, Dong, and Tang}]{yang2024trisampler}
Zhen Yang, Zhou Shao, Yuxiao Dong, and Jie Tang. 2024.
\newblock Trisampler: A better negative sampling principle for dense retrieval.
\newblock In \emph{Proceedings of the AAAI Conference on Artificial Intelligence}, volume~38, pages 9269--9277.

\bibitem[{Yu et~al.(2021)Yu, Liu, Xiong, Feng, and Liu}]{yu2021few}
Shi Yu, Zhenghao Liu, Chenyan Xiong, Tao Feng, and Zhiyuan Liu. 2021.
\newblock Few-shot conversational dense retrieval.
\newblock In \emph{Proceedings of the 44th International ACM SIGIR Conference on research and development in information retrieval}, pages 829--838.

\bibitem[{Yu et~al.(2022)Yu, Xiong, Sun, Zhang, and Overwijk}]{yu2022coco}
Yue Yu, Chenyan Xiong, Si~Sun, Chao Zhang, and Arnold Overwijk. 2022.
\newblock Coco-dr: Combating distribution shift in zero-shot dense retrieval with contrastive and distributionally robust learning.
\newblock In \emph{Proceedings of the 2022 Conference on Empirical Methods in Natural Language Processing}, pages 1462--1479.

\bibitem[{Yu et~al.(2024)Yu, Zhuang, Zhang, Meng, Ratner, Krishna, Shen, and Zhang}]{yu2024large}
Yue Yu, Yuchen Zhuang, Jieyu Zhang, Yu~Meng, Alexander~J Ratner, Ranjay Krishna, Jiaming Shen, and Chao Zhang. 2024.
\newblock Large language model as attributed training data generator: A tale of diversity and bias.
\newblock \emph{Advances in Neural Information Processing Systems}, 36.

\bibitem[{Zhan et~al.(2021)Zhan, Mao, Liu, Guo, Zhang, and Ma}]{zhan2021optimizing}
Jingtao Zhan, Jiaxin Mao, Yiqun Liu, Jiafeng Guo, Min Zhang, and Shaoping Ma. 2021.
\newblock Optimizing dense retrieval model training with hard negatives.
\newblock In \emph{Proceedings of the 44th International ACM SIGIR Conference on Research and Development in Information Retrieval}, pages 1503--1512.

\bibitem[{Zhou et~al.(2022)Zhou, Gong, Liu, Zhao, Shen, Dong, Lu, Majumder, Wen, and Duan}]{zhou2022simans}
Kun Zhou, Yeyun Gong, Xiao Liu, Wayne~Xin Zhao, Yelong Shen, Anlei Dong, Jingwen Lu, Rangan Majumder, Ji-rong Wen, and Nan Duan. 2022.
\newblock \href {https://doi.org/10.18653/v1/2022.emnlp-industry.56} {{S}im{ANS}: Simple ambiguous negatives sampling for dense text retrieval}.
\newblock In \emph{Proceedings of the 2022 Conference on Empirical Methods in Natural Language Processing: Industry Track}, pages 548--559, Abu Dhabi, UAE. Association for Computational Linguistics.

\end{thebibliography}

\appendix

\clearpage

\noindent In the appendix, we present three sections: in Appendix~\ref{sec:Appendix Prompt}, we describe the prompt we used, in Appendix~\ref{sec:Appendix Experiment Details}, we provide additional experiment details for reference, and in Appendix~\ref{sec:Appendix Case}, we demonstrate several cases from our experiment for intuitive display.

\section{Prompt Display}~\label{sec:Appendix Prompt}
In this section, we present the prompt for generating negatives, as detailed in the following ``Prompt for Negative Generation'', along with the attributes specification for each dataset.

\begin{tcolorbox}[%`colback`=gray!10,
            colframe=gray,
            width=1\linewidth,
            arc=1mm, 
            auto outer arc,
            title={Prompt for Negative Generation},
            breakable,
            ]
            
    Assume you are an expert in \textcolor{orange}{{\{domain\_name\}}}, and there is a example with a ``user\_query'' and its related doc ``positive\_document''.  \\
    
    example: \{example\} \\
    
    \textcolor{blue}{\textit{\textbf{[Task Definition]}}} \\
    Your task is to write three hard negative samples in JSON format. The JSON object must contain the following keys: \\
    - ``reasoning'': a string, reasoning steps on how to generate three hard negative documents.\\
    - ``hard\_negative\_document\_1'': a string, a hard-negative document to the user query.\\
    - ``hard\_negative\_document\_2'': a string, a hard-negative document to the user query.\\
    - ``hard\_negative\_document\_3'': a string, a hard-negative document to the user query.\\

    \textcolor{blue}{\textit{\textbf{[Reasoning Definition]}}} \\
    - Write the inference process step by step in ``reasoning'', including how to associate from the ``user\_query'' and "positive\_document" to get the hard-negative documents.\\
    \textcolor{blue}{\textit{\textbf{[Hard Negatives Definition]}}} \\
    - All the hard negative documents should use similar keywords or topics as the ``positive\_document''. \\
    - All the hard negative documents appear to address the ``user\_query'' at first glance. However, subtly diverges in content or context such that it does not truly answer the query or meet the user's information need.\\
    - All the hard negative documents should be plausible and accurate documents, they should be diverse in topic, sources, and styles.\\
     \textcolor{blue}{\textit{\textbf{[Attributes Definition]}}} \\
    - All the negative documents should be in the education level of \textcolor{orange}{\{difficult\_level\}} to comprehend, and the length should be \textcolor{orange}{{\{length\}}} the ``positive\_document''. \\
    \textcolor{blue}{\textit{\textbf{[Format Definition]}}} \\
    - Your output must always be a JSON object only, do not explain yourself or output anything else.
    \end{tcolorbox}
    
For each dataset, we specify different attributes according to its content. During prompt generation, we randomly select one item from each attribute to compose the final prompt, as shown below:

\begin{tcolorbox}[%`colback`=gray!10,
            colframe=gray,
            width=1\linewidth,
            arc=1mm, 
            auto outer arc,
            title={Attributes Specification},
            breakable,
            ]

    \textcolor{red}{\# For SciFact}: \\
    domain\_name = [ "Epidemiology", "Public Health", "Virology", "Biostatistics", "Healthcare Policy", "Infectious Diseases", "Bioinformatics", "Medical Research", "Pharmacology" ] \\
    difficult\_level = [ "Foundational (Equivalent to Elementary and Middle School)" , "Intermediate (High School and Undergraduate)" ] \\
    length = ["approximately the same as", "nearly the same as"] \\

    \textcolor{red}{\# For FiQA}: \\
    domain\_name = ["Finance"] \\
    difficult\_level = [ "Advanced (Postgraduate and Beyond)" ] \\
    length = ["approximately the same as", "nearly the same as"] \\

    \textcolor{red}{\# For Quora}: \\
    domain\_name = ["General knowledge"] \\
    difficult\_level = [ "Foundational (Equivalent to Elementary and Middle School)",
                         "Intermediate (High School and Undergraduate)" ] \\
    length = ["approximately the same as", "nearly the same as"]

    \textcolor{red}{\# For HotpotQA}: \\
    domain\_name = ["General knowledge"] \\
    difficult\_level = [ "Foundational (Equivalent to Elementary and Middle School)",
                         "Intermediate (High School and Undergraduate)" ] \\
    length = ["approximately the same as", "nearly the same as"]

    \textcolor{red}{\# For MS Marco}: \\
    domain\_name = ["Question Answering"] \\
    difficult\_level = [ "Foundational (Equivalent to Elementary and Middle School)",
                         "Intermediate (High School and Undergraduate)" ] \\
    length = ["approximately the same as", "nearly the same as"]

    \end{tcolorbox}

\section{Experiment Supplementation} \label{sec:Appendix Experiment Details}
\subsection{Implementation Detail}
During negative sample generation, we use the GPT-4o with the temperature setting of 0.7. Our implementation is based on the architecture of FlagEmbedding\footnote{https://github.com/FlagOpen/FlagEmbedding} with 8 NVIDIA Tesla V100 GPUs. For all dense retrieval models, we set the maximum query input length to 64 and the maximum passage input length to 512. The training batch size is set to 4, with 5 epochs and a learning rate of 1e-5. For models, we use the following checkpoints from Huggingface: BGE as bge-large-en-v1.5\footnote{https://huggingface.co/BAAI/bge-large-en-v1.5}, E5 as multilingual-e5-base\footnote{https://huggingface.co/intfloat/multilingual-e5-base}, and COCO-DR as cocodr-base-msmarco\footnote{https://huggingface.co/OpenMatch/cocodr-base-msmarco/tree/main}, which are then used for downstream fine-tuning. For LLMs, we choose gte-Qwen2-7B-instruct\footnote{https://huggingface.co/Alibaba-NLP/gte-Qwen2-7B-instruct} and gte-Qwen2-1.5B-instruct\footnote{https://huggingface.co/Alibaba-NLP/gte-Qwen2-1.5B-instruct}, with LoRA at rank 16 for fine-tuning. For the implementation baselines of ADORE and SimANS, to ensure a fair comparison, we consistently substitute the dense retriever with BGE. Other details follow the original papers.As for results evaluation, we use MTEB~\cite{muennighoff2022mteb}.  

\subsection{{Cost of Synthetic Data Generation}} \label{sec:cost}
{For synthetic data generation, the statistics are shown in Table~\ref{tab:cost}, which display the number of generations and the approximate cost.}
\begin{table}[htbp]
  \centering
  \vspace{-2mm}
  \scalebox{0.7}{
    \begin{tabular}{cccccc}
    \toprule
    
        \textbf{} & \textbf{FIQA}  & \textbf{SciFact} & \textbf{Quora} &\textbf{HotpotQA} & \textbf{MS Marco} \\

    \midrule

        ~ Num. & 5.2K & 0.8K & 5.0K & 66.4K & 6.9K \\

        ~ Cost. & $\approx$40\$ & $\approx$2\$ & $\approx$30\$ & $\approx$420\$ & $\approx$55\$ \\

    \bottomrule
    
    \end{tabular}}
    \vspace{-2mm}
      \caption{{Cost of Generation: ``Num.'' denotes the number of generations for each instance, while ``Cost.'' represents the total generation cost.}}
      \vspace{-2mm}
      \label{tab:cost}
      \end{table}

\subsection{Simple Prompt} \label{sec:Simple Prompt}
We present the following simple prompt used in the experiments conducted in Section~\ref{sec:harder experiment}. Compared to the full prompt template, we exclude the processes of self-reflection and attribute specification, retaining only the definition of hard negatives.

\begin{tcolorbox}[%`colback`=gray!10,
colframe=gray,
width=1\linewidth,
arc=1mm, 
auto outer arc,
title={Simple Prompt for Negative Generation},breakable,]

Assume you are an expert in \{domain\_name\}, and there is a example with a ``user\_query'' and its related doc ``positive\_document''. \\
example:\{example\_q\_pd\}\\

Your task is to write three hard negative samples in JSON format. The JSON object must contain the following keys: \\
- ``hard\_negative\_document\_1'': a string, a hard-negative document to the user query.\\
- ``hard\_negative\_document\_2'': a string, a hard-negative document to the user query.\\
- ``hard\_negative\_document\_3'': a string, a hard-negative document to the user query.\\

- All the hard negative documents length should be approximately the same as the ``positive\_document''. \\
- Your output must always be a JSON object only, do not explain yourself or output anything else.
\end{tcolorbox}

\section{Case Study} \label{sec:Appendix Case}
In this section, we present cases to intuitively showcase different synthetic generation for each datasets.

\begin{tcolorbox}[%`colback`=gray!10,
colframe=gray,
width=1\linewidth,
arc=1mm, 
auto outer arc,
title={FIQA Case},
breakable,]

\textcolor{red}{\textbf{Query}}: Why should I choose a business checking account instead of a personal account?\\

\textcolor{red}{\textbf{Positive Document}}: Some benefits of having a business checking account ( versus a personal checking account ) are: The first 3 should be pretty easy to determine if they are important to you. \#4 is a little more abstract, though I see you have an LLC taxed as a sole proprietorship, and so I'm guessing protecting your personal assets may have been one of the driving reasons you formed the LLC in the first place. If so, ``following through'' with the business account is advised.\\

\textcolor{red}{\textbf{Reasoning}}: The user query is asking for reasons to choose a business checking account over a personal account. The positive document gives a clear answer to this. For the hard negative examples, I will choose topics that are related to finance and banking, but do not directly answer the user's question. They will seem relevant at first glance due to the use of similar keywords and topics, but upon closer examination, it will be clear that they do not provide the specific information the user is seeking. The documents will be diverse in sources, ranging from general finance advice, to information about savings accounts, and finally a document detailing the process of opening a business account without highlighting the benefits over a personal one.\\

\textcolor{red}{\textbf{Hard Negative Document 1}}: When it comes to managing your finances, it's important to make sure you're making the right choices for your situation. A checking account is a type of deposit account that allows for numerous withdrawals and unlimited deposits. On the other hand, a savings account is a deposit account held at a bank or other financial institution that provides principal security and a modest interest rate. Depending on your financial goals and needs, you might choose one over the other.\\

\textcolor{red}{\textbf{Hard Negative Document 2}}: A savings account is a deposit account at a bank or other financial institution that provides principal security and a modest interest rate. Unlike a checking account, a savings account may have restrictions on the number of withdrawals or transfers you can make each month. It’s also possible to incur a fee if you go over the limit. As the name implies, savings accounts are designed to help you save money, so they typically come with higher interest rates than checking accounts.\\

\textcolor{red}{\textbf{Hard Negative Document 3}}: Opening a business checking account is a relatively straightforward process. You'll need to provide some basic information about your business, including its legal name, physical address, and federal tax ID number. You'll also need to provide your personal identification, such as a driver's license or passport. Once you've gathered all of this information, you can go to the bank or credit union where you want to open your account. They'll walk you through the process and help you get your account set up.
\end{tcolorbox}

\begin{tcolorbox}[%`colback`=gray!10,
colframe=gray,
width=1\linewidth,
arc=1mm, 
auto outer arc,
title={Quora Case},
breakable,]

\textcolor{red}{\textbf{Query}}: How do I get rid of mosquitoes bites quickly?\\

\textcolor{red}{\textbf{Positive Document}}: How do I get rid of mosquito bite marks?\\

\textcolor{red}{\textbf{Reasoning}}: To generate the hard negative documents, I would start by identifying the central theme in the user query and positive document, which is 'mosquito bites'. For the first hard negative document, I would use the same keywords to discuss a topic that's closely related but does not answer the user's question about getting rid of mosquito bite marks quickly. The second hard negative document would be focused on a different aspect of mosquito bites, such as prevention, rather than treatment. The third hard negative document would be about a different type of insect bite, using similar language and keywords but diverging in the specific type of insect discussed.\\

\textcolor{red}{\textbf{Hard Negative Document 1}}: What are some common symptoms of mosquito bites?\\

\textcolor{red}{\textbf{Hard Negative Document 2}}: How can I prevent mosquito bites when I go camping?\\

\textcolor{red}{\textbf{Hard Negative Document 3}}: What should I do if I get a spider bite?
    \end{tcolorbox}

\begin{tcolorbox}[%`colback`=gray!10,
colframe=gray,
width=1\linewidth,
arc=1mm, 
auto outer arc,
title={SciFact Case},
breakable,]

\textcolor{red}{\textbf{Query}}: A single nucleotide variant the gene DGKK is strongly associated with increased risk of hypospadias.\\

\textcolor{red}{\textbf{Positive Document}}: Hypospadias is a common congenital malformation of the male external genitalia. We performed a genome-wide association study using pooled DNA from 436 individuals
with hypospadias (cases) and 494 controls of European descent and selected the highest ranked SNPs for individual genotyping in the discovery sample, an additional Dutch sample of 133 cases
and their parents, and a Swedish series of 266 cases and 402 controls. Individual genotyping of two SNPs (rs1934179 and rs7063116) in DGKK, encoding diacylglycerol kinase, produced compelling evidence for association with hypospadias in the discovery sample (allele-specific odds ratio
(OR) = 2.5, P = 2.5 × 10-11 and OR = 2.3, P = 2.9 × 10-9, respectively) and in the Dutch (OR = 
3.9, P = 2.4 × 10-5 and OR = 3.8, P = 3.4 × 10-5) and Swedish (OR = 2.5, P = 2.6 × 10-8 and OR
= 2.2, P = 2.7 × 10-6) replication samples. Expression studies showed expression of DGKK in
preputial tissue of cases and controls, which was lower in carriers of the risk allele of
rs1934179 (P = 0.047). We propose DGKK as a major risk gene for hypospadias.\\

\textcolor{red}{\textbf{Reasoning}}: To create hard negative documents, I first analyzed the 'user query' which focuses on the association between a single nucleotide variant in the gene DGKK and an increased risk of hypospadias. The 'positive document' provides detailed evidence of this association, including statistical data from a genome-wide association study and individual genotyping that supports DGKK as a major risk gene for hypospadias. To generate hard negative documents, I ensured that each document would: 1) Use similar keywords or be related to the topics of genetics, congenital malformations, or specifically hypospadias; 2) Seemingly address the user query at first glance by discussing relevant genetic or medical research; 3) Subtly diverge in content or context, such as focusing on different genes, diseases, or aspects of genetic research that do not truly answer the query about the association between DGKK and hypospadias. Each hard negative document is crafted to be plausible and accurate, sourced from fictitious but believable studies or information, and maintains an advanced educational level for comprehension.\\

\textcolor{red}{\textbf{Hard Negative Document 1}}: Recent studies have identified a significant correlation between variants in the WNT4 gene and the development of Mullerian duct anomalies in females. A comprehensive review of genomic data from 500 female patients with congenital reproductive anomalies and a control group of 520 healthy females revealed that specific SNPs in the WNT4 gene were associated with a higher incidence of these conditions. While this research underscores the importance of genetic factors in congenital reproductive malformations, it does not directly relate to the association of DGKK with hypospadias in males, despite its focus on congenital malformations.\\

\textcolor{red}{\textbf{Hard Negative Document 2}}: In an exploration of the genetic basis of male infertility, a team of researchers conducted a meta-analysis of genetic association studies targeting genes involved in testicular function. Among the genes examined, FSHR and AR were highlighted for their roles in spermatogenesis. Although this research provides valuable insights into the genetic determinants of male reproductive health, it diverges from the specific inquiry into the DGKK gene's association with hypospadias, focusing instead on different genes and a different aspect of male reproductive health.\\

\textcolor{red}{\textbf{Hard Negative Document 3}}: A recent investigation into the genetic underpinnings of pediatric urinary tract infections (UTIs) has identified several candidate genes that may influence susceptibility to these infections. Utilizing a cohort of 300 pediatric patients with recurrent UTIs and 350 controls, the study found no significant association between the DGKK gene and UTIs. However, it did highlight the potential role of other genes in the immune response to urinary pathogens. While this document discusses genetic susceptibility to a specific condition affecting the urinary tract, it does not address the link between DGKK gene variants and hypospadias.

\end{tcolorbox}

\begin{tcolorbox}[%`colback`=gray!10,
colframe=gray,
width=1\linewidth,
arc=1mm, 
auto outer arc,
title={HotpotQA Case},
breakable,]

\textcolor{red}{\textbf{Query}}: What type of study does University of Cape Town and Queen's University have in common?
\\

\textcolor{red}{\textbf{Positive Document}}: The University of Cape Town (UCT) is a public research university located in Cape Town in the Western Cape province of South Africa. UCT was founded in 1829 as the South African College making it the oldest higher education institute in South Africa, it is jointly the oldest university in South Africa and the oldest extant university in Sub-Saharan Africa alongside Stellenbosch University which received full university status on the same day in 1918. UCT is the highest-ranked African university in the QS World University Rankings, the Times Higher Education World University Rankings, and the Academic Ranking of World Universities, and its Law and Commerce Faculties are consistently placed among the hundred best internationally. The language of instruction is English.
\\

\textcolor{red}{\textbf{Reasoning}}: To generate hard negative documents, we start by understanding the user query and the positive document. The user query is asking for the common type of study between the University of Cape Town (UCT) and Queen's University. The positive document primarily provides details about UCT, including its history, ranking, and faculties. To create hard negatives, we need documents that seem relevant by mentioning universities, research, or related topics, but do not actually answer the query about the common type of study between UCT and Queen's University. The hard negatives should be on similar topics, use similar keywords, and be plausible, but they must diverge subtly from the actual query, creating an illusion of relevance without fulfilling the information need.
\\

\textcolor{red}{\textbf{Hard Negative Document 1}}: Queen's University, located in Kingston, Ontario, Canada, is a prestigious public research university established in 1841. Known for its comprehensive programs in engineering, business, and science, Queen's University consistently ranks among the top Canadian universities. The university's commitment to research excellence is evident through its numerous research centers and institutes. The language of instruction at Queen's University is English.
\\

\textcolor{red}{\textbf{Hard Negative Document 2}}: The University of Oxford is a collegiate research university in Oxford, England. It is the oldest university in the English-speaking world, with evidence of teaching dating back to 1096. The university is renowned for its rigorous academic programs and has produced numerous notable alumni, including 28 British Prime Ministers. Oxford's research output is among the highest in the world, and it is frequently ranked as one of the top universities globally.\\

\textcolor{red}{\textbf{Hard Negative Document 3}}: The University of Melbourne, founded in 1853, is a leading public research university located in Melbourne, Australia. It is known for its strong emphasis on research and teaching, offering a wide range of undergraduate and postgraduate programs. The University of Melbourne is highly regarded in fields such as medicine, engineering, and the arts. It is consistently ranked among the top universities in Australia and the world.
\end{tcolorbox}

\begin{tcolorbox}[%`colback`=gray!10,
colframe=gray,
width=1\linewidth,
arc=1mm, 
auto outer arc,
title={MS Marco Case},
breakable,]

\textcolor{red}{\textbf{Query}}: what is bergamot flavoring?
\\

\textcolor{red}{\textbf{Positive Document}}: Citrus bergamia, the bergamot orange, is a fragrant fruit the size of an orange, with a yellow color similar to a lemon. Genetic research into the ancestral origins of extant citrus cultivars found bergamot orange to be a probable hybrid of Citrus limetta and Citrus aurantium.
\\

\textcolor{red}{\textbf{Reasoning}}: To generate hard negative documents, we start by analyzing the user query 'what is bergamot flavoring' and the positive document provided. The positive document describes the bergamot orange, its size, color, and genetic origins. The key elements to focus on for hard negatives include the citrus nature, fruit characteristics, and potential hybrid origins. However, the hard negatives should not directly address bergamot flavoring. For the first document, we can discuss the uses of bergamot in perfumery, which shares the citrus and aromatic elements but does not explain the flavoring specifically. The second document can describe another citrus fruit with similar characteristics, such as the pomelo, thus using relevant terms but diverging on the actual subject. The third document can discuss the use of citrus in a culinary context, such as in desserts, again not directly addressing bergamot flavoring but appearing relevant initially.
\\

\textcolor{red}{\textbf{Hard Negative Document 1}}: Bergamot is widely used in the perfume industry due to its refreshing and citrusy aroma. Its essential oils are extracted from the peel and utilized in many fragrances. The unique scent of bergamot is often described as sweet and floral, yet slightly spicy, making it a popular choice for men's and women's perfumes alike.
\\

\textcolor{red}{\textbf{Hard Negative Document 2}}: The pomelo, scientifically known as Citrus maxima, is a large citrus fruit similar in appearance to a grapefruit. It has a thick, greenish-yellow rind and a sweet, mildly tart pulp. The pomelo is believed to be one of the ancestors of the grapefruit and is often enjoyed fresh or in salads.\\

\textcolor{red}{\textbf{Hard Negative Document 3}}: Citrus fruits such as lemons, limes, and oranges are frequently used in desserts for their tangy flavor. Lemon zest and juice can be found in cakes, tarts, and sorbets, providing a fresh and zesty taste that complements sweet ingredients. Citrus is a versatile addition to many recipes, enhancing both flavor and aroma.
\end{tcolorbox}

\label{sec:appendix}

\end{document}